\documentclass[reprint, aps,superscriptaddress]{revtex4-2}
\usepackage{graphicx}% Include figure files
\usepackage{dcolumn}% Align table columns on decimal point
\usepackage{epsfig,latexsym,cancel,amssymb,amsmath,bm}
\usepackage{bm}% bold math
\newcommand{\vv}{v}
\newcommand{\cc}{c}
\newcommand{\aaa}{f}
\newcommand{\zz}{V}
\newcommand{\ff}{\rho}
\newcommand{\nn}{N}

\begin{document}
%\templatetype{pnasresearcharticle} % Choose template 
% {pnasresearcharticle} = Template for a two-column research article
% {pnasmathematics} %= Template for a one-column mathematics article
% {pnasinvited} %= Template for a PNAS invited submission
\title{Optimal intercellular competition in senescence and cancer}

% Use letters for affiliations, numbers to show equal authorship (if applicable) and to indicate the corresponding author
\author{Thomas C. T. Michaels} 
\affiliation{Department of Physics and Astronomy, Institute for the Physics of Living Systems, University College London, Gower Street, London WC1E 6BT, UK}
\affiliation{Medical Research Council Laboratory for Molecular Cell Biology, University College London, Gower Street, London, WC1E 6BT, UK}
\author{L. Mahadevan}
\affiliation{Engineering and Applied Sciences, Physics, and Organismic and Evolutionary Biology, Harvard University, Cambridge, MA 02138, USA}

\begin{abstract}
Effective multicellularity requires both cooperation and competition between constituent cells. Cooperation involves sacrificing individual fitness in favor of that of the community, but excessive cooperation makes the community susceptible to senescence and aging. Competition eliminates unfit senescent cells via natural selection and thus slows down aging, but excessive competition makes the community susceptible to cheaters, as exemplified by cancer and cancer-like phenomena. These observations suggest that an optimal level of intercellular competition in a multicellular organism maximizes organismal vitality by delaying the inevitability of aging. We quantify this idea using a statistical mechanical framework that leads to a generalized replicator dynamical system for the population of cells that change their vitality and cooperation due to somatic mutations that make them susceptible to aging and/or cancer.  By accounting for the cost of cooperation and strength of competition in a minimal setting, we show that our model predicts an optimal value of competition that maximizes vitality and delays the inevitability of senescence or cancer. 
\end{abstract}

\maketitle

The evolution of multicellularity is linked to the advantages of collective physiology and behavior absent in unicellular life which include, but are not limited to \cite{strassmann2010social,aktipis2015cancer,michod2000darwinian,pfeiffer2003evolutionary}:  division of labor,  adaptation to varied environments, efficient use of resources,  creation and maintenance of extracellular environmental niches, the collective  inhibition of cell proliferation and programmed cell death.  All these benefits of multicellular life require cooperation - the coordinated orchestration of functions that are essential for the development and maintenance of a complex organism. Cooperation, however, comes with a cost to individual cells that need to invest a part of their resources into traits that contribute positively to organismal vitality but reduce individual cell fitness \cite{baillon2014reflections,michod1996cooperation,claveria2016cell}. 

Cooperation in multicellar organisms creates interdependence among cells, which in turn leads to damage accumulation, gradual decay and aging. Indeed, it has been shown that many complex systems with multiple connected components (including both biologically evolved organisms and artificially engineered systems) in general experience aging as a result of the interdependence between the components \cite{harman1981aging,vural2014aging,sun2020optimal,taneja2016dynamical,farrell2016network,mitnitski2017aging}. To ameliorate the consequences of degradation induced by aging in both evolved and engineered systems requires continuous maintenance and repair. In multicellular organisms, a particular form of maintenance, controlled cell proliferation, increases the risk of accumulating deleterious heritable somatic mutations that cause the progressive decline of cellular function and, eventually, an irreversible arrest of cell growth, i.e. cellular senescence \cite{nelson2017intercellular,vijg2000somatic,campisi2013aging}. The accumulation of senescent cells results in a progressive loss of organismal vitality and a number of aging-related pathologies \cite{van2014role,campisi2013aging}. 

In multicellular organisms with renewable tissues, senescent cells can be eliminated through natural selection as a result of the reduced fitness of senescent cells compared to healthy cells \cite{wodarz2007effect,biteau2010lifespan,chalmers2012cell,baillon2014reflections,campisi2013aging,van2014role}. The resulting intercellular competition serves to increase organismic vitality at a cost associated with proliferation-driven renewal. The potential for the breakdown of cellular cooperation driven by excessive proliferation can lead to inappropriate cell survival, resource monopolization, abnormal cell differentiation, or degradation of the extracellular environment, which are considered hallmarks of cancer \cite{hanahan2011hallmarks,tenen2003disruption,aktipis2015cancer,Gil2016,hausser2020tumour}. This is a form of cheating that emerges in a competitive environment because uncooperative cancerous cells enjoy a higher fitness relative to cooperative healthy cells and have an advantage in selection \cite{baillon2014reflections}. Then individually uncooperative cells can thrive (transiently) in a competitive environment, with deleterious consequences for the (long-term) collective vitality of the organism \cite{Goodell1199}. 

Thus intercellular competition in multicellular organisms is a double-edged sword: without competition, multicellularity is susceptible to senescence, while too much competition can lead to cheating and cancer or cancer-like phenomena \cite{aktipis2015cancer,Gil2016}. Recently, an elegant study \cite{nelson2017intercellular} builds on this idea suggesting that senescence and cancer are an inevitable consequence of the dilemma posed by competition that cannot be too weak or too strong, and spawned a series of commentaries on the generality of the conclusions \cite{wagner2017power,cheong2018multicellular,mitteldorf2018questioning}.  Some  questions that naturally arise in this context include the possibility of a minimal analytic framework that might help to uncover the essence of the arguments, while also posing the problem of whether there is an optimal level of competition that maximizes organismal vitality by controlling or delaying senescence without succumbing to cancer? 

Here we attempt to answer this question in terms of an approach based on a probabilistic master equation. We show that this leads to an analytically tractable mathematical model for the dynamics of multicellular aging in terms of a modified form of generalized replicator dynamics. Our solvable model reveals the fundamental factors controlling the optimal level of intercellular competition in terms of  two parameters that characterize the base fitness and vitality in the system. Our solution reveals the fundamental biophysical factors controlling the  level of intercellular competition and predict an optimal value of this parameter that maximizes system vitality by delaying the inevitability of senescence. By providing a minimal statistical mechanical framework for the study of the collective vitality of a system, our results may motivate the rational design of   strategies for delaying aging in biological, technical or social systems, where processes similar to those considered here are at play.

\begin{figure}
\begin{center}
\includegraphics[width=0.4\textwidth]{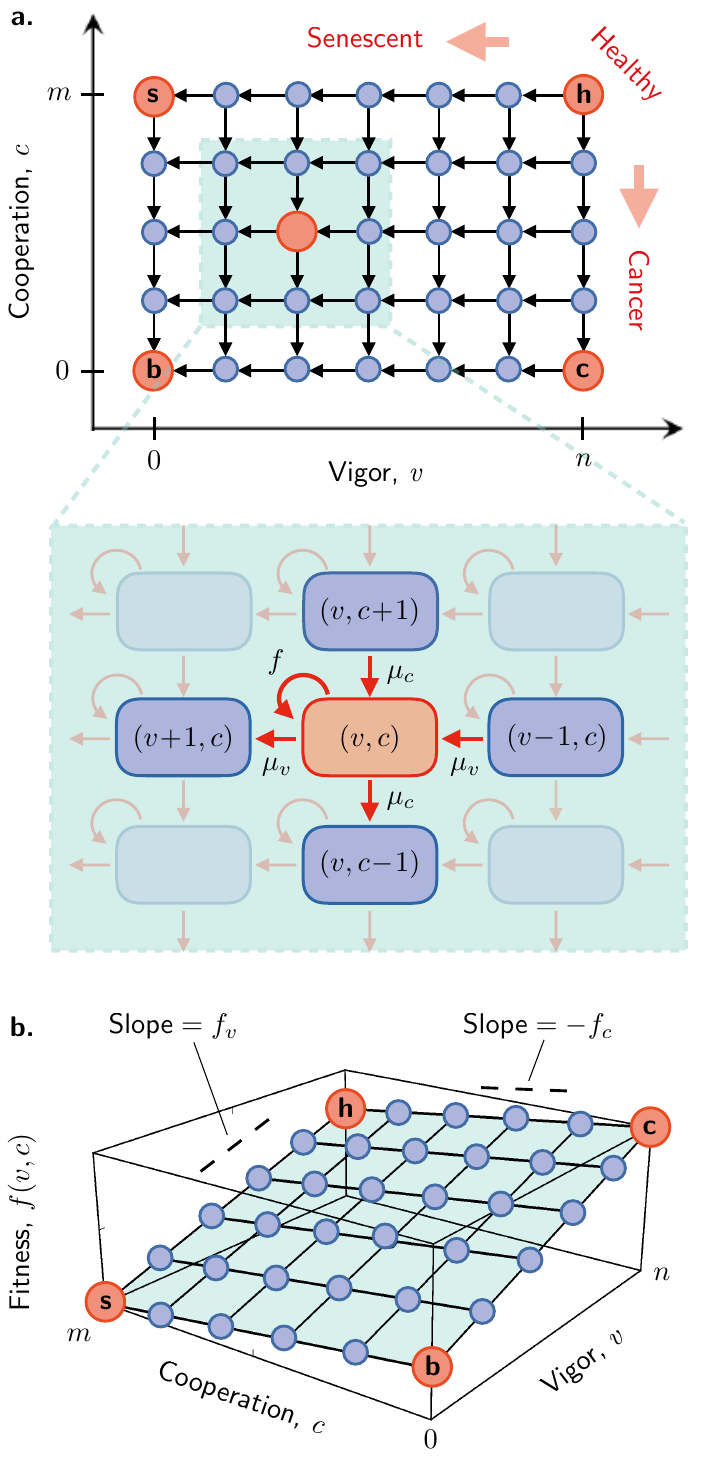}
\end{center}
\caption{\textbf{\textsf{Master equation for multicellular aging.}} \textbf{\textsf{a.}} Cell vigor $\vv$ and cooperation $\cc$ define a two-dimensional coordinate system $(\vv,\cc)$ of cell types. Progressive loss of vigor corresponds to cell senescence, while progressive loss of cooperation corresponds to cancer. %\textbf{\textsf{b.}} 
The dynamics of the population $\nn(\vv,\cc,t)$ of cells of type $(\vv,\cc)$ at time $t$ are described in terms of transition rates between different states $(\vv,\cc)\to (\vv',\cc')$. The figure illustrates the different fluxes away from and into state $(\vv,\cc)$: (i) each cell type $(\vv,\cc)$ proliferates with state-dependent rate  $\aaa(\vv,\cc)$; (ii) mutations correspond to transitions that lower vigor ($\vv\to \vv-1$, with rate $\mu_{\vv}$) or cooperation ($\cc\to \cc-1$, with rate $\mu_{\cc}$). 
\textbf{\textsf{b.}} Fitness (competition) landscape for different cell types $(\vv,\cc)$ for $k=0.3$. Senescent cells are non-competitive with $f(0,m)=0$. Healthy cells are moderately competitive with $\aaa(n,m) = \aaa_0(1-k)$, while cancer cells are the most competitive species with $\aaa(n,0)=\aaa_0$. In the model it is important that $\partial \aaa/\partial \vv >0$ but $\partial \aaa/\partial \cc <0$, i.e.~the fitness landscape is tilted in favour of cancerous types. This condition ensures that loss of cooperation in the form of cheating corresponds to an advantage in selection.
\label{fig1}}
\end{figure}

 \section*{Dynamics of multicellular aging}

To describe the dynamics of multicellular aging, we use a master equation approach in the space of cell types, as an equivalent to the Price equation formalism \cite{queller2017fundamental,frank2012natural,okasha2006evolution} (see SI Sec.~S1 for a master equation formulation of the Price equation). Following a previous study~\cite{nelson2017intercellular}, we classify cell types in terms of two traits: vigor $\vv$ and cooperation $\cc$. Vigor $\vv$ is used as a general measure of cellular resources or function, e.g. metabolic activity. Cooperation $\cc$ describes the fraction of resources that a cell devotes in activities that favor the functioning of the multicellular organism, including controlling homeostasis or maintaining the extracellular infrastructure. Since it is measured in terms of the fraction of the vigor, we expect it to be an intensive variable. Vigor and cooperation define a coordinate system of discrete cell types $(\vv,\cc)$; in the simplest setting, we assume that $\vv$ and $\cc$ take discrete values in the range $0\leq \vv \leq n$ and $0\leq \cc \leq m$ (Fig.~\ref{fig1}(a)) (a formulation using continuous values of these variables does not lead to results that are qualitatively different - see SI Sec.~S2 and Fig.~S1). Cells with high vigor ($\vv=m$) and high cooperation ($\cc=m$) are `healthy' ({\it{h}}), cell types that have lost vigor ($\vv=0$) are `senescent' ({\it{s}}), while cells with $\cc=0$ are `cancerous' ({\it{c}}) \cite{nelson2017intercellular}. Cells in state $(\vv=0,\cc=0)$ are both senescent and cancerous ({\it{b}}). Cells with intermediate values of $\vv$ and $\cc$ represent types that are not fully degraded; this situation mimics the fact that multiple mutations are necessary to induce senescence or cancer \cite{promislow1998mutation}.

\subsection*{Cell populations} 

To quantify the time evolution of the (average) population $\nn(\vv,\cc,t)$ of cells in state $(\vv,\cc)$ at time $t$ we use a (mean field) master equation approach \cite{krapivsky2010kinetic}. This captures spontaneous transitions between states in the coordinate system $(\vv,\cc)$. In our system transitions occur as a result of two effects (Fig.~\ref{fig1}(a)): (i) cell proliferation (i.e.~intercellular competition) at a rate $f(v,c)$ and (ii) somatic mutations, which are permanent changes of cell genotype that lead to a gradual decline of vigor or cooperation phenotypes. %not just c mutations (which are beneficial to the cell) but also v mutations (which are deleterious) 
In the following we shall use the terms fitness and competition interchangeably to denote the proliferation rate $f(v,c)$. We consider somatic mutations that affect only one of the two traits $\vv$ or $\cc$ in single steps with (state-dependent) rates $\mu_{\vv} \vv$, respectively, $\mu_{\cc}\cc$ \cite{nelson2017intercellular}, but note that our framework can in principle be generalised to account for more complex transitions. Then the master equation describing the time evolution of $\nn(\vv,\cc,t)$ is given by:
\begin{align}
\label{multiplesomat1}
\frac{\partial \nn(\vv,\cc,t)}{\partial t} & = \aaa(\vv,\cc) \, \nn(\vv,\cc,t)  \\ \nonumber
&  + \mu_{\vv}(\vv+1)\, \nn(\vv+1,\cc,t) - \mu_{\vv}\, \vv\,   \nn(\vv,\cc,t)  \\ \nonumber
&  +  \mu_{\cc}(\cc+1)\, \nn(\vv,\cc+1,t)- \mu_{\cc}\, \cc\,   \nn(\vv,\cc,t)\, .
\end{align}
The first term on the right-hand side of \eqref{multiplesomat1} describes the proliferation of $\nn(\vv,\cc,t)$ with growth rate $\aaa(\vv,\cc)$. The remaining terms in \eqref{multiplesomat1} describe the effect of somatic mutations by means of reaction fluxes away from and into the different cell types, i.e. $\nn(\vv,\cc,t)$ decreases due to transitions $\vv\to \vv-1$ and $\cc\to \cc-1$; conversely, $\nn(\vv,\cc,t)$ increases through transitions $\vv+1 \to \vv$ or $\cc+1 \to \cc$.

\subsection*{Cell fractions}  

Since it is more useful to consider the dynamics of population fractions rather than  the time evolution of various cell populations, we define the fraction of cells in state $(\vv,\cc)$ at time $t$ as: \begin{equation}
    \ff(\vv,\cc,t)=\frac{\nn(\vv,\cc,t)}{\nn(t)}, \quad \nn(t)=\sum_{\vv,\cc} \nn(\vv,\cc,t)\, ,
\end{equation}  
where $\nn(t)$  is the total population  of cells. We can then use \eqref{multiplesomat1} and derive a dynamic equation for $\ff(\vv,\cc,t)$ (see SI Sec.~S1) as: 
\begin{subequations}\label{multiplesomat}
\begin{align}
\frac{\partial \ff(\vv,\cc,t)}{\partial t} & = \big(\aaa(\vv,\cc) - \overline{\aaa}(t)\big)\, \ff(\vv,\cc,t) \nonumber \\
&  + \mu_{\vv}(\vv+1)\, \ff(\vv+1,\cc,t) - \mu_{\vv}\, \vv \, \ff(\vv,\cc,t) \nonumber \\ 
&  +  \mu_{\cc}(\cc+1)\, \ff(\vv,\cc+1,t)- \mu_{\cc}\, \cc\,  \ff(\vv,\cc,t)\, ,
\end{align}
where 
\begin{equation}
 \overline{\aaa}(t) = \sum_{\vv,\cc} \aaa(\vv,\cc)\,  \ff(\vv,\cc,t)
\end{equation}
\end{subequations}
is the (time-dependent) average competition. Compared to \eqref{multiplesomat1}, the key difference of \eqref{multiplesomat} lies in the first line, which describes the effect of selection: cell types with $\aaa  > \overline{\aaa}$ will be enriched by selection (fitness is higher than average), while those with $\aaa  < \overline{\aaa}$ will decrease in frequency over time (fitness lower than average). \eqref{multiplesomat} is thus an extension of the replicator equation \cite{cressman2014replicator} that accounts for mutation fluxes changing cell type. Note that while the master equation for cell populations \eqref{multiplesomat1} is  linear, the equation for cell fractions \eqref{multiplesomat} is non linear. Also note that the master equation is a deterministic equation describing the average population or fraction of cells of a certain type, while $v$ and $c$ are stochastic variables, capturing stochasticity in the appearance of mutations or in cell proliferation. Alternative methods commonly used in population genetics include PDE-based approaches such as the Fokker-Planck formalism \cite{gomez2020mutation}, which follow from the master equation through Taylor-series expansion in $v$ and $c$; here we stay with the ME approach for generality.

We choose the mutation rates $\mu_{vv},\mu_{cc}$ to be linearly dependent on $\vv$ and $\cc$ respectively for simplicity. This choice guarantees that in the absence of selection the average vigor $\overline{\vv}(t) = \sum_{v,c} \vv\, \rho(v,c,t) \propto e^{-\mu_{\vv} t}$ and cooperation $\overline{\cc}(t) = \sum_{v,c} \cc\, \rho(v,c,t) \propto e^{-\mu_{\cc} t}$ decay exponentially with rates $\mu_{\vv}$, respectively, $\mu_{\cc}$, and naturally ensures that mutation transitions do not reduce $v$ or $c$ below zero (boundary condition at $\vv=0$, $\cc=0$). 

\begin{figure}
\begin{center}
\includegraphics[width=0.48\textwidth]{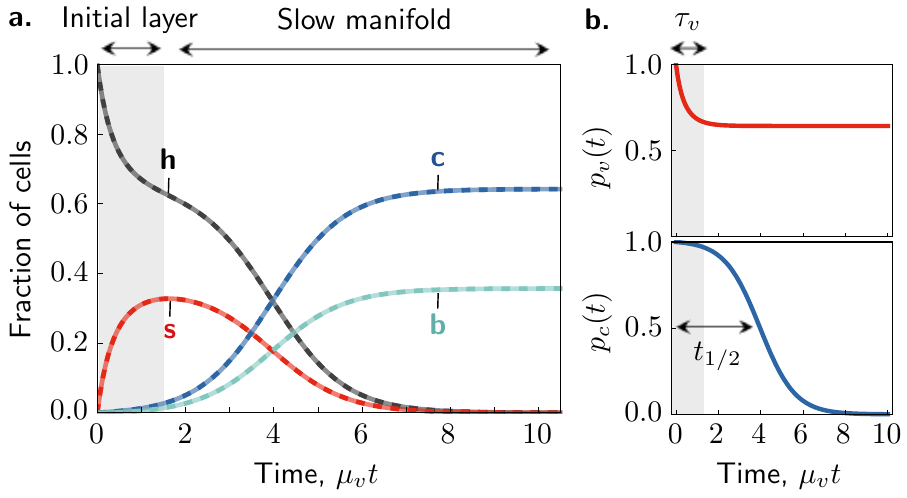}
\end{center}
\caption{\textbf{\textsf{Dynamics of multicellular aging.}} 
\textbf{\textsf{a.}} Time evolution of fractions of healthy ({\it{h}}), senescent ({\it{s}}), cancerous ({\it{c}}) and both senescent and cancerous ({\it{b}}) cells obtained as solution to the master equation (\ref{multiplesomat}) for $k=0.3$, $\mu_{\vv}=10^{-3}$, $\mu_{\cc}=10^{-5}$, $\aaa=0.004$ and $n=m=1$ (four-state model). Solid lines indicate the numerical solution to \eqref{multiplesomat}, while dashed lines indicate the exact analytical solution (given by \eqref{maselsol}) for initial conditions associated with $(v,c) = (1,1), (0,1), (1,0), (0,0)$) corresponding to the four-state model (see texxt for details). 
\textbf{\textsf{b.}} Time evolution of the probabilities $p_v(t)$ and $p_c(t)$, defined in \eqref{maselsol}, and graphical representation of the timescales $\tau_{\vv}$ and $t_{1/2}$.
 \label{fig2}}
\end{figure}

\begin{figure*}
\begin{center}
\includegraphics[width=0.85\textwidth]{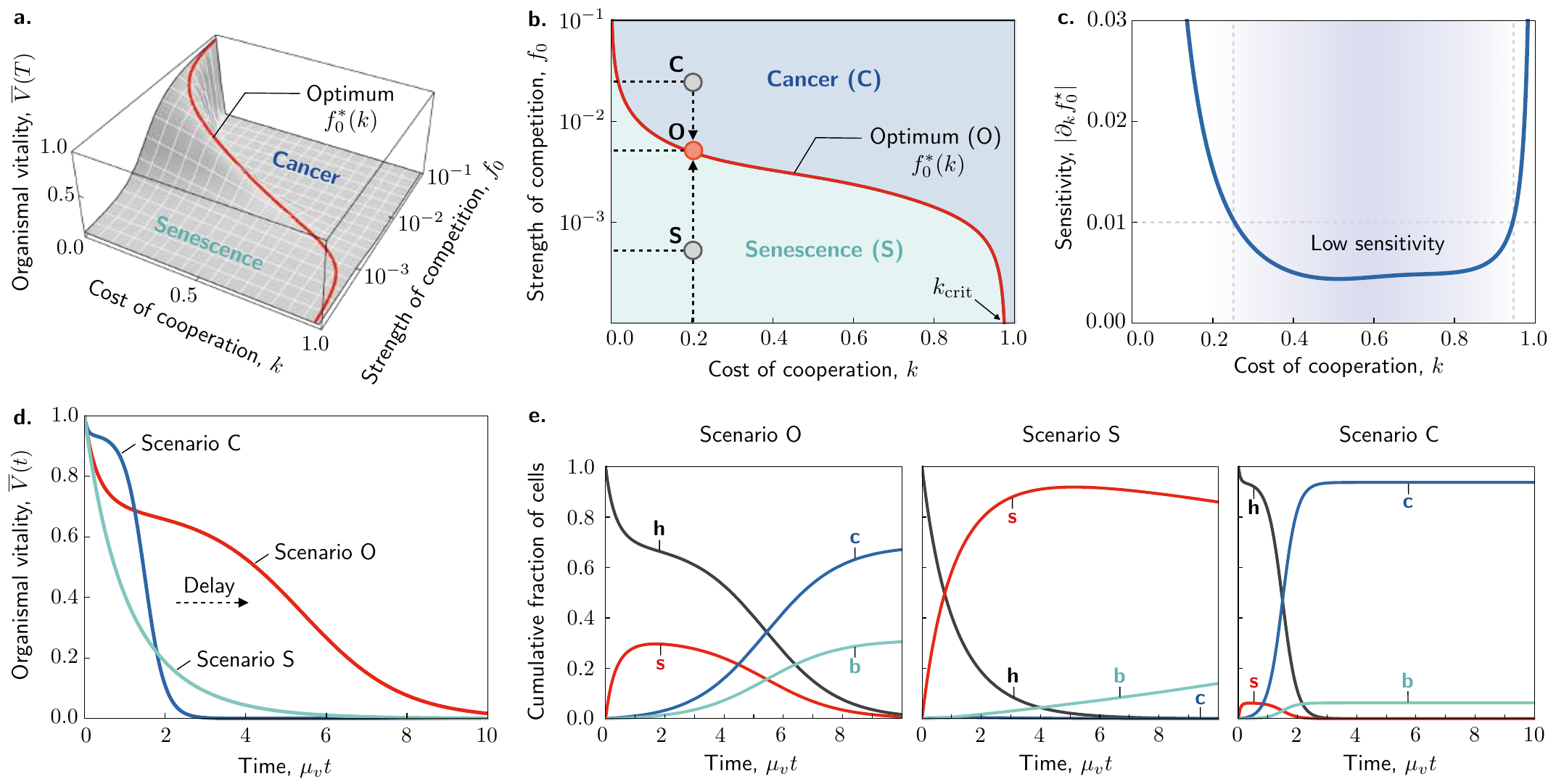}
\end{center}
\caption{\textbf{\textsf{Organismal vitality and optimal competition.}} 
\textbf{\textsf{a.}} Average organismal vitality $\overline{\zz}(T)$ at observation time $T=3\mu_v^{-1}$ plotted against the cost of cooperation $k$ and the strength of cellular competition $\aaa_0$. The solid line indicates the position of the optimum line $f_0^\star(k)$. The plot is obtained by numerical integration of \eqref{multiplesomat} using the same parameters as in Fig.~\ref{fig2}(a). 
\textbf{\textsf{b.}} Optimal competition $\aaa_0^\star(k)$ in terms of $k$ (this corresponds to a top view of \ref{fig3}(a)). Points on the optimal line $f_0^\star(k)$ (scenario O) balance the accumulation of senescent cells and development of cancer. Points in the ($k$,$\aaa_0$)-diagram that deviate from the optimal line (scenarios S and C) correspond to systems that are either dominated by senescence or cancer.
\textbf{\textsf{c.}} Sensitivity of optimal conditions, defined as $|\partial \aaa_0^\star/\partial k|$. The plot is obtained by numerically evaluating the derivative of the optimal line $f_0^\star(k)$ of \ref{fig3}(b) along $k$. Regions of low sensitivity facilitate a robust approach to optimal conditions.
\textbf{\textsf{d.}} Time evolution of average organismal vitality $\overline{V}(t)$ for the scenarios O, S and C defined in \ref{fig3}(b). \textbf{\textsf{e.}} Fractions of healthy, senescent and cancerous cell types ({\it{h}}, {\it{s}}, {\it{c}}, and {\it{b}}) corresponding to scenarios O (left), S (middle) and C (right). The optimal scenario O balances senescent cell accumulation and cancer cell proliferation.
\label{fig3}}
\end{figure*}

 \subsection*{Time evolution of cell fractions}

To solve the master equation \eqref{multiplesomat} we need to specify a closure relation for $\aaa(\vv,\cc)$. This procedure is analogous to path or contextual analysis of the Price equation \cite{li1975path}, where fitness, or other parameters entering the Price equation, are partitioned into separate causes through regression equations \cite{queller2017fundamental,frank2012natural,okasha2006evolution}. In the simplest setting we assume a linear relationship for $\aaa(\vv,\cc)$ (corresponding to a one-term Taylor expansion) by imposing two minimal requirements: $f$ should {\it{increase}} with increasing $\vv$, and {\it{decrease}} with increasing $\cc$, i.e.
\begin{equation}\label{pder}
\frac{\partial \aaa}{\partial \vv}>0 ,  \frac{\partial \aaa}{\partial \cc}<0\, .
\end{equation}
In Fig.~\ref{fig1}(b), we depict the relative fitness of the different cell types: senescent cells (s) have the lowest fitness with $f=0$, cancer cells (c) have the highest fitness $\aaa_0$, while healthy cells have an intermediate fitness $f=f_0(1-k)$, where $k\in [0,1]$ is the cost of cooperation. A linear fitness function  that satisfies these requirements can then be written as $f(\vv,\cc) = f_0 k + f_{\vv} \, \vv - f_{\cc}\, \cc$, where $f_{\vv}=f_0(1-k)/n$ and $f_{\cc}=-f_0k/m$, such that $f(n,m)=f_0(1-k)$, $f(n,0)=f_0,f(0,m)=0$ and $f(0,0)=f_0k$. We see that cells that have higher vigor or invest a smaller fraction of their resources on sustaining the organism have an advantage in selection compared to less vigorous or more cooperative types.

With the simple choice in \eqref{pder}, the master equation \eqref{multiplesomat} can now be solved using the method of generating functions \cite{krapivsky2010kinetic} to obtain an exact analytical solution for the evolution of the population fractions of cells (details in Materials and Methods), yielding 
\begin{subequations}\label{maselsol}
\begin{align}
\ff(t,\vv,\cc) & = \textrm{Bin}\big(\vv,n,p_\vv(t)\big)\, \textrm{Bin}\big(\cc,m,p_\cc(t)\big) \, .
\end{align}
Here $\textrm{Bin}(v,n,p) = {{n}\choose{\vv}} \, p^v (1-p)^{n-v}$ denotes the binomial distribution and the functions $p_\vv(t)$ and $p_\cc(t)$ are given by
\begin{align}
p_\vv(t) & = \frac{f_\vv-\mu_\vv}{f_\vv - \mu_\vv \, e^{-t/\tau_v}} \, , \\
p_\cc(t) & = \frac{f_\cc+\mu_\cc}{f_\cc + \mu_\cc \, e^{t/\tau_c}} \, ,
\end{align}
\end{subequations}
where $\tau_\vv = 1/(f_\vv-\mu_{\vv})$ and $\tau_\cc=1/(f_\cc+\mu_\cc)$ are the two natural timescales that control the dynamics in our model. 
The analytical solution reveals that $\ff(t,\vv,\cc)$ is the product of two independent binomial distributions in vigor and cooperation spaces, with $p_\vv(t)$ and $p_\cc(t)$ representing the (time-dependent) probabilities of cells having one unit of vigor or cooperation, respectively.

Fig.~\ref{fig2}(a) illustrates the time evolution of the various cell fractions for $n=m=1$. This corresponds to a four-state model where cells can be in one of four states at time $t=0$: healthy ({\it{h}}, $(v,c)=(1,1)$), senescent ({\it{s}}, $(v,c)=(0,1)$), cancerous ({\it{c}}, $(v,c)=(1,0)$), and both senescent and cancerous ({\it{b}}, $(v,c)=(0,0)$). We also assume that $\mu_{\vv} \gg \mu_{\cc}$ capturing the observation that only about one percent of human genes contributes to cancer risk \cite{futreal2004census}, so that ~$\mu_{\cc}/\mu_{\vv} \simeq 10^{-2}$

The separation of timescales between senescence-causing and cancer-causing mutations implied by $\mu_{\vv} \gg \mu_{\cc}$ causes the resulting complex aging dynamics to display two stages of kinetics, as reflected by the distinct timescales controlling the time evolution of the probabilities $p_\vv$ and $p_\cc$ (Fig.~\ref{fig2}(b)). Initially there is a phase of characteristic timescale $\tau_\vv $ where a buildup of senescent cells is observed as a result of accumulation of senescence-causing mutations. In this rapidly varying initial phase, a rapid pre-equilibrium is established between healthy and senescent cells, with the fraction of healthy cells pre-equilibrating at $p_{\vv}(\infty)^n = [1-\mu_{\vv}/(kf_{0})]^n$ and the fraction of senescent cells approaching the maximal value $[\mu_{\vv}/(kf_{0})]^n$. This pre-equilibrium reflects the interplay between selection $kf_0$ and mutation forces $\mu_{\vv}$; stronger competition eliminates senescent cells more effectively via natural selection hence increasing the pre-equilibrium fraction of healthy cells.
During this stage of dynamics, degradation events causing cancer are negligible at leading order (i.e.~$p_\cc\simeq 1$) and the fraction of cancerous cells stays close to zero. After this rapid initial phase, the solution develops into a second, slower phase (corresponding to the slow manifold \cite{hinch_1991}), where senescent cells are slowly removed by the combined action of selection and cancer-causing mutations and the fraction of cancerous types is seen to increase with time. Cancerous cells display sigmoidal kinetics, increasing slowly initially then more rapidly and eventually reaching a plateau. At the end of this second phase of dynamics, the fractions of healthy and senescent cells approach zero, while {\it{c}} and {\it{b}} cells form an equilibrium with fractions $ [1-\mu_{\vv}/(kf_{0})]^n$ and $[\mu_{\vv}/(kf_{0})]^n$, respectively. A measure of the characteristic timescale of cancer development is given by the time needed for $p_{\cc}(t)$ to drop by a factor of two (the half-life), yielding $t_{1/2} = \tau_{\cc} \ln(2+f_{\cc}/\mu_{\cc})$, i.e. it is directly proportional to $\tau_{\cc}$ with a logarithmic correction that depends on the ratio $f_c/\mu_c$. The half-life $t_{1/2}$ increases with decreasing $\aaa_{\cc}$ or $\mu_{\cc}$. In particular, for $\mu_{\cc}\to 0$ we find $t_{1/2}\to \infty$, suggesting that in the absence of cancer-causing mutations, intercellular competition is able to maintain the pre-equilibrium between healthy and senescent cells indefinitely. In reality, for any value of $\mu_{\cc}>0$, however small, the final ($t \to \infty$) state of the system consists of an equilibrium between {\it{c}} and {\it{b}}  cells and no healthy cells, in line with the idea that above some threshold age, the mortality rate monotonically increases with time \cite{nelson2017intercellular}. % of inevitable multicellular aging \cite{nelson2017intercellular}.

\subsection*{Organismal vitality and optimal competition}

Having an understanding of the population dynamics using our master equation formalism, we now turn to question the relative importance of the various ``microscopic'' (cell-level) processes and their contribution to ``macroscopic'' (organism-level) observables, and in particular the strength of intercellular competition. To do so, a natural definition of a minimal model of average organismal vitality is the population-weighted average
\begin{equation}\label{avz}
\overline{\zz}(t) \equiv \sum_{\vv,\cc} \zz(\vv,\cc)\, \ff(\vv,\cc,t)\, ,
\end{equation}
where $\zz(\vv,\cc)$ describes the contribution of cell type $(\vv,\cc)$ to the vitality of the whole organism. Since vitality requires cells to be both vigorous and cooperative, a simple choice for $\zz(\vv,\cc)$ is:
\begin{equation}\label{eneq}
\zz(v,c) = \zz_0\, \vv\, \cc,
\end{equation}
where $\vv \cc$ measures the amount of resources devoted to cooperative activities and $V_0$ is a pre-factor that sets the units of vitality. This multiplicative choice is guided by the thought that for a multicellular organism to be vital, we need cells to be both vigorous and cooperative. Since the mean of the binomial distribution $\textrm{Bin}(v,n,p)$ is $\overline{\vv} =np$, for this choice of $\zz(v,c)$, the average vitality takes the simple form $ \overline{\zz}(t) = \zz_0\, \overline{\vv}(t)\, \overline{\cc}(t) =\zz_0\, n\,m\,  p_{\vv}(t)\, p_{\cc}(t)$. 

Using the analytical solution for  $\ff(\vv,\cc,t)$ given by [\ref{maselsol}] we optimize $\overline{\zz}(T)$ with respect to the strength  of intercellular competition $\aaa_0$, for times $T \gg 1/f_0, 1/\mu_v$, i.e. much larger than the timescales of proliferation and mutation (the dependence of the results on the choices of these parameters is considered later). Figure \ref{fig3}(a) shows a plot of the average organismal vitality $\overline{\zz}(T)$ at the observation time $T=3\mu_v^{-1}$ as a function of the cost of cooperation $k$ and the strength of competition  $\aaa_0$. We see that $\overline{\zz}$ has a non-monotonic behaviour with $\aaa_0$.  In Fig.~\ref{fig3}(b), we show the distinct maximum of vitality as a function of the base level of competition characterized by $f_0$ and the cost of cooperation $k$, characterized by an optimal curve $\aaa_0^\star(k)$ (see SI Sec.~S3) for asymptotic analytical expressions for this optimal line $f_0^\star(k)$). Inspection of Fig.~\ref{fig3}(b) shows that when $k=0$ the optimal strength of competition diverges, i.e.~$f_0^\star \to \infty$. This is intuitive since for $k=0$ there is no cost for cooperation. Cancer and healthy cells have therefore the same fitness, which implies that cheating gives cancer cells no advantage in selection. 
As we increase the cost of cooperation, $k$, the optimal competition $f_0^\star(k)$ gradually decays indicating that the system cannot tolerate high levels of intercellular competition when the cost of cooperation becomes large. Eventually $f_0^\star(k)$ becomes zero at a critical value $k=k_{\rm{crit}}$ and then stays identically zero for larger values $k>k_{\rm{crit}}$. In this limit, the fitness of healthy cells is so low that any non-zero amount of intercellular competition will result in a dominance of cancer. The critical $k_{\rm{crit}}$ depends on a combination of the rates of mutation (see SI Sec.~S3 and Fig.~S3 for the exact analytical expression for $k_{\rm{crit}}$). When $\mu_v/\mu_c\to 0$ the critical $k_{\rm{crit}}\to 0$, while for $\mu_v/\mu_c \to \infty$ we have $k_{\rm{crit}}\to 1$. When $\mu_v/\mu_c\to 1$ we find $k_{\rm{crit}}\to 1/2$.
%When $k\to 1$ the optimal strength of competition tends to $f_0^\star \to 0$. In this limit healthy cells have vanishing fitness. Any non-zero amount of intercellular competition will therefore result in a dominance of cancer. 
We note that the level of optimal competition also depends on the rates of mutations causing senescence $\mu_{\vv}$ and cancer $\mu_{\cc}$ and on the observation time $T$ (see Fig.~S2). Increasing $\mu_{\vv}$ requires a higher optimal level of competition to counteract the stronger tendency of senescent cell accumulation, while increasing $\mu_{\cc}$ causes $f_0^\star$ to decrease since the system is more susceptible to cheating. Larger times $T$ require a lower level of intercellular competition, reflecting a balance between shorter-term advantage of cheating and longer-term detrimental effects of cheating on organismal vitality. 

The sensitivity of the value of the optimal competition is defined by the derivative of the optimal line $\aaa_0^\star(k)$ with respect to the cost of cooperation $k$, i.e.~$|\partial f_0^\star/\partial k|$.  High sensitivity means that optimal competition is easily affected by changes in $k$, i.e.~small fluctuations in $k$ will push the system away from the optimum. In contrast, low sensitivity implies that optimal competition is relatively insensitive to the choice of $k$, which facilitates a robust approach to the optimum. In Fig.~\ref{fig3}(c)), we show that to optimize vitality while maintaining low sensitivity a plausible solution is to operate in the regime of low to moderate $k$, as this choice yields a robust strategy with relatively high vitality.

To understand the dynamics the populations as a function of the level of competition, we note that in Fig.~\ref{fig3}(b), the  optimal curve $\aaa_0^\star(k)$ divides $(k,\aaa_0)$-space into two separate regions, which we term the {\it{senescent}} region and the {\it{cancerous}} region. As shown in Fig.~\ref{fig3}(d), choosing the parameters $(k,\aaa_0)$ along the optimal curve (scenario O) delays the loss of organismal vitality, corresponds to a balance between senescence cell accumulation and cancer proliferation. Compared to this situation, when parameters are chosen in the senescent region (scenario S) the loss of organismal vitality is dominated by accumulation of senescent cells (Fig.~\ref{fig3}(e)), while in the cancerous region (scenario C) loss of vitality is driven by the proliferation of cancer cells (Fig.~\ref{fig3}(e)).   
%Another important parameter is the derivative of optimal competition with respect to the cost of cooperation $k$, i.e.~$\kappa^*=|\partial f_0^\star/\partial k|$, which we term sensitivity. Sensitivity controls the achievability of optimal conditions. Optimal competition is very sensitive in regions where $\kappa^*$ is large, which occurs when $k$ is close to 0 or 1. Here, small fluctuations in $k$ will push you away from the optimum. By contrast, sensitivity is low for intermediate values of $k$, which facilitates a robust approach to the optimum. These results suggest that optimal conditions correspond to low cost of cooperation, where optimal organismal vitality is large and sensitivity is low.

\section*{Discussion}

In this study, we have proposed a model for aging in terms of the dynamics of a multicellular population characterized by its vigor and cooperation, and a fitness parameter that is a function of these variables. Using a master-equation based framework, we derived a replicator-like dynamical equation accounting for fluxes due to mutations that lead to cancer and senescence. Assuming a minimal closure relation for the fitness that is linear in the vigor and cooperation, we are led to an analytic solution for the evolution of the relative fraction of healthy, senescent and cancerous populations while accounting for the interplay between competition and cooperation. Using our model and a simple choice of organismal vitality that is multiplicative in the vigor and cooperation, we then probed the optimal level of intercellular competition that maximizes organismal vitality, balancing the cost of cooperation and the strength of competition. 

It might be interesting to interpret our result of an optimal level of competition in terms of Parrondo's paradox as highlighted in a commentary \cite{cheong2018multicellular}, whereby combining losing strategies may lead to a winning strategy \cite{harmer1999losing,harmer2001brownian,jian2014beyond}. This idea has been invoked in a range of biological systems to explain adaptation in areas ranging from genetics to ecology \cite{cheong2019paradoxical,reed2007two,tan2017nomadic}. In the context of multicellular aging, too little competition is a losing strategy because it leads to senescent cell accumulation. On the other hand, too much competition is also a losing strategy because even if senescent cells are eliminated it leads to cheater-cell proliferation in the long term. Following the logic of Parrondo's paradox, we see that a fruitful approach to delay multicellular aging combines both (losing) strategies to achieve a fine balance that delays senescence and staves off cancer. In this optimal scenario,  competition is strategically reduced when cheater cells over-proliferate and increased back when senescent cells over-accumulate. A similar strategy may be realised with coexisting subpopulations of uncompetitive and competitive cells if the ratio between the subpopulations is strategically changed in response to the changing environment. 

Our framework may be extended in multiple ways. Effective multicellularity not only requires cooperation between cells, but also mechanisms for suppressing conflict that result from it. This effect may be accounted for by introducing a feedback between strength of competition and the current system state. To investigate the impact of drugs that clear senescent or cancer cells \cite{baker2016naturally}, we can envisage coupling the master replicator equation with dynamic equations for an inhibitor that removes deleterious cell types. Generalizing our results could thus provide a framework for interrogating the effect of strategies to combat aging and cancer dynamics and how to optimize them.  Furthermore, our model considers detrimental mutations reducing vigor or cooperation as irreversible. Even though proliferation arrest in senescence cells is essentially irreversible, certain biological manipulations, including inactivation of specific tumor suppressor genes, can reverse senescence \cite{beausejour2003reversal}. The latter scenario could be studied by making transitions between cell types in our model reversible, and more generally by including separate dynamic equations modeling changes of $v$ and $c$ over evolutionary timescales to model the effect of tumor suppressor genes. While we have limited ourselves here to the balance between cooperation and competition during the later stages of life associated with aging, similar questions arise during the developmental stages of organisms when proliferation rates of cells are high and thus susceptible to mutational errors \cite{nichols2022cell}. Understanding the dynamic regulation (and misregulation) of intercellular competition in early development might also be amenable to our approach.  

While we have limited ourselves to studying the role of discrete mutations, it is not hard to generalize our framework to the case of continuously varying vigor and cooperation. This leads to a Boltzmann-like equation for the dynamics of cellular populations (see SI Sec.~S2). Finally, it is worth noting the natural appearance of extensive variables such as vigor (and vitality) and intensive variables such as cooperation (and competition), suggesting natural analogies to (non-equilibrium) thermodynamics. The relation between these variables and various forms of closure relations will be explored in a separate study.

{We acknowledge support from the UCL Institute for the Physics of Living Systems (TCTM), the Swiss National Science Foundation (TCTM), Peterhouse, Cambridge (TCTM), the Simons Foundation (LM) and the Henri Seydoux Fund (LM).}

 %s\bibliographystyle{pnas}
 \bibliography{References}

\end{document}